\documentclass[summary]{URSIGASS2021}

\usepackage{amssymb}
\usepackage{amsmath}

\usepackage[numbers]{natbib}	

\title{RadioLuna:\\
A Penetrometer Deployed Network For Lunar Radio Science Below 2 MHz}

\author{T. M. Eubanks*, W. P. Blase}

\affiliation{%
{Space Initiatives Inc\\
Newport, Virginia, USA\\
tme@space-initiatives.com\\}
}

\begin{document}

\maketitle

\begin{abstract}
The radio environment of the Moon at low frequencies, particularly in 
lunar polar regions and the permanently shadowed regions (PSR) found there, is relatively poorly explored and may contain some novel features. In addition, these areas of the Moon, shielded from the natural and artificial emissions of the Earth, and the natural radio emissions of the Sun and the other planets, are near-ideal locations for radio astronomy observations in the last unexplored region of the electromagnetic frequency spectrum. We are developing a low-mass RadioLuna radio science precursor mission to deploy an
interferometric array on the lunar surface using 
Space Initiatives ``Mote'' penetrometers. The current RadioLuna default mission would be an array of 10 - 12 penetrators  deployed on the floor of Shackleton crater in the PSR 4 km below the crater rim, where it would be shielded from all terrestrial, solar and planetary interference. 
At the present level of understanding of the low frequency lunar radio environment there will be a tight coupling between advances in technology and advances in science, and RadioLuna can be expected to lead to improvements in both lunar radio science and lunar communication techniques. 
\end{abstract}

\section{Introduction}

The radio bands below the radio plasma frequency of the Earth's ionosphere ($\sim$ 2 - 20 MHz, depending on conditions) represent the last major astronomically unexplored portion of the electromagnetic spectrum. This region of the radio spectrum is effectively inaccessible from the Earth's surface, and there have been plans and missions to 
explore it astronomically  since Radio-Astronomy-Explorer-1 and 2 in the early 1970s \cite{Alexander-et-al-1975-a,Basart-et-al-1997-a}. It was rapidly realized that the natural radio emissions of the Earth and Sun can be overpowering in this spectral band, leading to mission proposals for radio telescopes and arrays on the lunar far side \cite{Jester-Falcke-2009-a,Burns-et-al-2012-a,Mimoun-et-al-2012-a}, and to the recent 
successful Netherlands Chinese Low Frequency Explorer (NCLE) launched to the Earth-Moon Lagrange Point 2 as a part of the Chinese Chang'E-4 Lunar mission in 2018 \cite{Bentum-et-al-2020-a}.

Ballistic penetrators will enhance lunar science by allowing for the rapid initial scientific investigation of difficult to reach terrain such as the polar PSRs, volcanic vents, crater central peaks and lava tube skylights, together with the rapid placement of instrument arrays in regions of interest \cite{Smrekar-et-al-1999-a,Eubanks-et-al-2020-a}. Here, we discuss an application of penetrator technology for radio science \cite{Eubanks-et-al-2020-c}.

\section{RadioLuna}

RadioLuna will be a penetrometer deployed network of small receivers and transmitters with a total deployed mass of order 20 kg.
RadioLuna will characterize the lunar radio and charged particle environment in two poorly explored radio bands, at frequencies below 30 KHz and between 0.06 - 2 MHz, neither of which are accessible from the Earth.
The primary RadioLuna scientific goals are
\begin{enumerate}
	\item To passively monitor the low frequency (2 MHz and below) radio environment on the Moon.
	\item To search for natural  astronomical radio sources in that band.
	\item To determine the ability to obtain reflections from the lunar wake (at night time) and the Earth's magnetotail (near Full Moon), to monitor the physical parameters of these plasma structures and determine their usefulness of these reflections for low bit rate emergency communications.
	\item To explore the ability to send ground waves on the Moon at $\sim$1 MHz, for communication, and to observe the bulk dielectric  constants of the regolith along the transmission path.
	\item To determine the ability of supporting communications by direct transmission through the lunar regolith at low frequencies. 
	\item To passively search for the Earth's Auroral Kilometric Radiation in the 60 - 500 KHz band, and to determine if these strong natural radio emissions can be received through kilometers of lunar rock and regolith.
	\item To monitor the plasma passing over the array during and after lunar landings and meteorite impacts. 
\end{enumerate}
Some of these goals can be accomplished by passive monitoring by the array, others will require transmission from one Mote and reception by other Motes, either in the same network or (after multiple deployments) in other networks. 

In a 1 month nominal mission, RadioLuna should expect to address all of these scientific and engineering goals. although of course mission goals would benefit from longer mission durations. 
 Subsequent missions would build on the findings from the early missions; having two or more RadioLuna arrays simultaneously operating on the Moon would make it possible to 
to send MHz ground waves and reflected kHz radio waves  over long distances, and possibly even to conduct very long baseline interferometry of astronomical sources in the 1 MHz band. 

\section{Mote Ballistic Penetrators}
\begin{figure}[!ht]
\begin{center}
\includegraphics[scale=0.20]{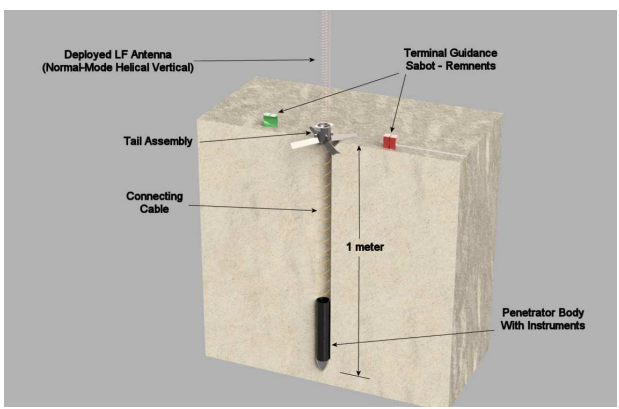}
\end{center}
\caption{A Mote penetrator after a nominal 1 meter deployment into the lunar regolith.
The electronics and most of the scientific payload would be carried in the  penetrator itself
and would be automatically carried 1 to 2 meters into the lunar regolith, while for RadioLuna the radio antennas (monopoles and magnetic coil antennas) would be deployed by the tail section onto the lunar surface. 
}
\label{fig:penetrator-in-regolith}
\end{figure}

Space Initiatives Inc (SII) has developed small ``Mote'' ballistic penetrators  to provide commercial support for robotic and crewed operations on and near the Moon.
The $\sim$1.5 kg Mote penetrators will have on-board processing, communications and sensors, and can be deployed from a CubeSat deployer by a dedicated mission or carried by a lander proceeding  to the lunar surface under NASA's Commercial Lunar Payload Services (CLPS) or Artemis programs. After deployment, Motes fall ballistically, impacting the surface at up to 300 m/s and, as shown in Figure \ref{fig:penetrator-in-regolith}, should penetrate 1 meter or more into typical lunar regolith, resulting in a a sensor array spread, in a nominal mission, over $\sim$1 km of the lunar surface.

\section{Deploying into Shackleton Crater}
\begin{figure}[!ht]
\begin{center}
\includegraphics[scale=0.625]{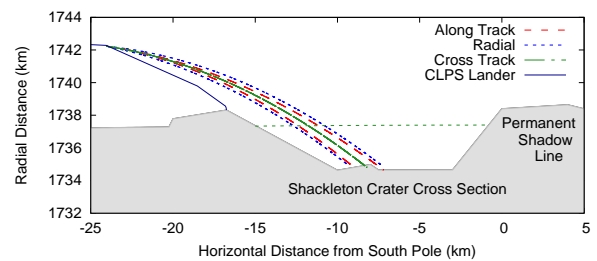}
\end{center}
\caption{Deployment of Mote Penetrators into the Shackleton Crater PSR. 
The Motes are assumed to be deployed from a CLPS Lander early enough in the landing sequence to proceed into the center of the PSR, while the CLPS lander proceeds to soft land at the crater rim. The CLPS landing profile shown here was developed from discussions with CLPS providers. While the actual landing profile will depending on the CLPS provider, in any realistic profile the lander will touchdown at least one and possibly several minutes after the Motes are on the surface. At no time will the penetrators come close to the lander after deployment. 
}
\label{fig:Drop_orbits_6}
\end{figure}

Ballistic penetrators will enable the rapid deployment of instruments into the most difficult lunar terrains, including the permanently shadowed regions. Figure \ref{fig:Drop_orbits_6} shows the deployment of Motes into Shackleton crater as part of a landing on the rim of the crater. The Motes are deployed 24 km downrange and  5 km above the mean lunar surface and take $\sim$78 seconds to reach the crater floor, $\sim$2.8 km below the mean lunar surface.
At the time of their landing, the lander itself would still be well above the surface of the Crater rim and would be able to observe IR emissions from the gas plumes emitted by surface volatiles vaporized by Mote impacts.  The horizontal spread of the Mote's landing sites, due to deployment separation velocities of $\le$10 m s$^{-1}$, is roughly a km  both along and transverse to their trajectory.

\section{The Plasma Environment at the Lunar Poles}

\begin{figure}[!ht]
\begin{center}
\includegraphics[scale=0.625]{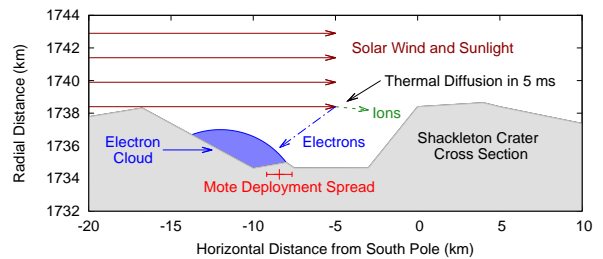}
\end{center}
\caption{Shackleton Crater partially filled with a non-neutral electron cloud. Shackleton Crater has a diameter of $\sim$20 km, which the supersonic solar wind, with a typical velocity of $\sim$400 km s$^{-1}$, would pass in about 50 ms. Due to equipartition of energy the thermal velocities of the constituents of the solar wind vary widely; at a typical temperature of 11 eV the electron acoustic velocity is $\sim$1400 km s$^{-1}$, while the proton ion acoustic velocity is only $\sim$32 km s$^{-1}$
\citep{Rhodes-Farrell-2019-a,Rhodes-Farrell-2019-b}. The electrons will diffuse into the crater interior in as little as 5 msec, while the ions cannot diffuse all the way into the depth of the crater in the crater crossing time, and so will presumably form back currents from the far wall of the crater. This could cause a non-neutral electron cloud inside the crater, leading to 
negative surface potentials exceeding  in the PSR \citep{Farrell-et-al-2020-a}.
The penetrator deployment  indicated here is that shown in Figure \ref{fig:Drop_orbits_6} and would 
allow observation of this process.}
\label{fig:Shackleton_4}
\end{figure}

Table \ref{table:regimes} briefly describes the complicated and time-variable plasma environment near the lunar surface. The Moon is in the complicated plasma dynamics  of the Earth's magnetotail roughly 25\% of the time, and the remainder of the time will be in the supersonic solar wind \cite{Bhardwaj-et-al-2015-a}. In the solar wind  
wake the e$^{-}$ density is substantially decreased, and the plasma can become non-neutral as the thermal velocities of the solar wind electrons 
are higher than the bulk wind velocity, while the thermal ion velocities are substantially lower.
Charge-stratified plasmas are thus likely to form in shadowed lunar craters \cite{Rhodes-Farrell-2019-b}, and possibly also in the lunar wake, in both cases on a scale that does not allow for laboratory simulation on Earth. These non-neutral electron clouds could produce significant charging of astronauts and their equipment in shadowed regions\cite{Rhodes-Farrell-2019-a} and must be better characterized to understand this safety hazard. Figure \ref{fig:Shackleton_4} indicates how charge-separation may  fill shadowed craters with non-neutral electron clouds, which may support a variety of electrostatic and cyclotron waves and a very complicated plasma environment
\cite{Zimmerman-et-al-2012-a}. It will be difficult to directly sample this environment from orbit (satellites cannot fly inside most of the shadowed lunar craters), but a RadioLuna array in a PSR could be used observe from inside the shadowed region, observing PSR plasma wave phenomena from the surface  in real time.

\begin{table}
        \begin{center}
        \begin{tabular}{l c r}
                \hline
                Region &  e$^{-}$ Density & Plasma \\
                       & \# cm$^{-3}$ & Frequency \\
                \hline
               Solar Wind &  $\sim$7 & $\sim$ 24 kHz \\
               UV-Disassociated e$^{-}$ & $\sim$600 & $\sim$ 2 MHz \\
               Earth Magnetotail & $\sim$0.3  & $\sim$1 kHz  \\
               Shadowed Craters & $\sim$0.01 & ?  \\
               Lunar Wake & $\sim$0.01  & $\sim$30 Hz  \\
               \hline
        \end{tabular}
        \end{center}
\caption{Typical Plasma Regions Near the Moon (conditions are subject to large variations). The shadowed craters (and possibly the wake) may contain non-neutral plasmas with a zero plasma frequency. References: \cite{Rhodes-Farrell-2019-b,Jensen-Vilas-2011-a,Manning-2008-a,Halekas-et-al-2005-a}. }
\label{table:regimes}
\end{table}

The thermal speed of electrons in the solar wind is faster than the bulk velocity, which should allow for shadowed craters at the lunar poles to fill with non-neutral electron clouds, which may support a variety of electrostatic and cyclotron waves and a very complicated plasma environment in those regions
\cite{Zimmerman-et-al-2012-a}. It will be difficult to directly sample this environment (satellites cannot fly inside most of the shadowed lunar craters), and they are much larger than any laboratory plasma, but RadioLuna Motes could be used to passively or actively monitor frequencies between 30 Hz and 30 KHz across or inside a shadowed lunar crater, allowing for mapping of crater wave electron densities in the non-neutral plasma. This use will require deployment of a RadioLuna network in a specific location, monitoring at specific times, or both.

\section{Lunar Communications at Very Low Frequencies}

The very low frequency lunar environment is relatively unexplored in general on the lunar surface. Manning \cite{Manning-2008-a} suggested that the photo-emitted electrons at the lunar surface would allow for long range ground-wave communications over the lunar day-time surface at f $\sim$ 1 MHz; this has never been tested on the lunar surface. These surface waves would of course sample the entire regolith dilectric propertoes and the photoemission density along the raypath. If communication could be established at $\sim$ 1 MHz between stations at the two lunar poles it 
could be possible to determine the bulk dielectric properties of large areas of the Moon, and test this form of long range communications. 

At very low frequencies, roughly at or below the solar wind plasma frequency of $\sim$24 kHz, 
``ionospheric'' radar may be possible using reflections off of the lunar wake, which would provide a novel means of monitoring conditions in the wake. Such reflections may also be useful for communications; even very low bit rate global communications would find use on the Moon, both as a means of distributing alerts and for emergency communications.

\section{Deep Space Astronomy}

In the 0.06 to 2 MHz band the largest sources of interference in cislunar space are the terrestrial AKR, type III solar bursts, and Jupiter-Io emissions, all of which would be shielded in an appropriate PSR \cite{Jester-Falcke-2009-a}. In general at low frequencies the largest radio source from outside the solar system is the galactic synchrotron emission background, which is strongest in the galactic plane. The flux density of this background declines at frequencies $\lesssim$ 2 MHz, as the brightness temperature flattens out. Although it is not clear if there are any sufficiently bright extragalatic radio sources sufficiently bright for detection by RadioLuna, it should be possible to make a low resolution image of the galactic background radiation.



\bibliographystyle{ieeetr}

\end{document}